\documentclass{article}
\usepackage{graphicx}
\usepackage[english]{babel}

\usepackage{amsmath}
\usepackage{hyperref}
\usepackage{color}
\usepackage{cite}

        \let\i = \iota

\newcommand{\beq}{\begin{equation}}
\newcommand{\eeq}{\end{equation}}
\newcommand{\bea}{\begin{eqnarray}}
\newcommand{\eea}{\end{eqnarray}}
\newcommand{\beqa}{\begin{eqnarray}}
\newcommand{\eeqa}{\end{eqnarray}}

\definecolor{red}{rgb}{1,0,0}

\def\be{\begin{equation}}
\def\ee{\end{equation}}
\numberwithin{equation}{section}

\title{
\includegraphics[width=0.35\textwidth]{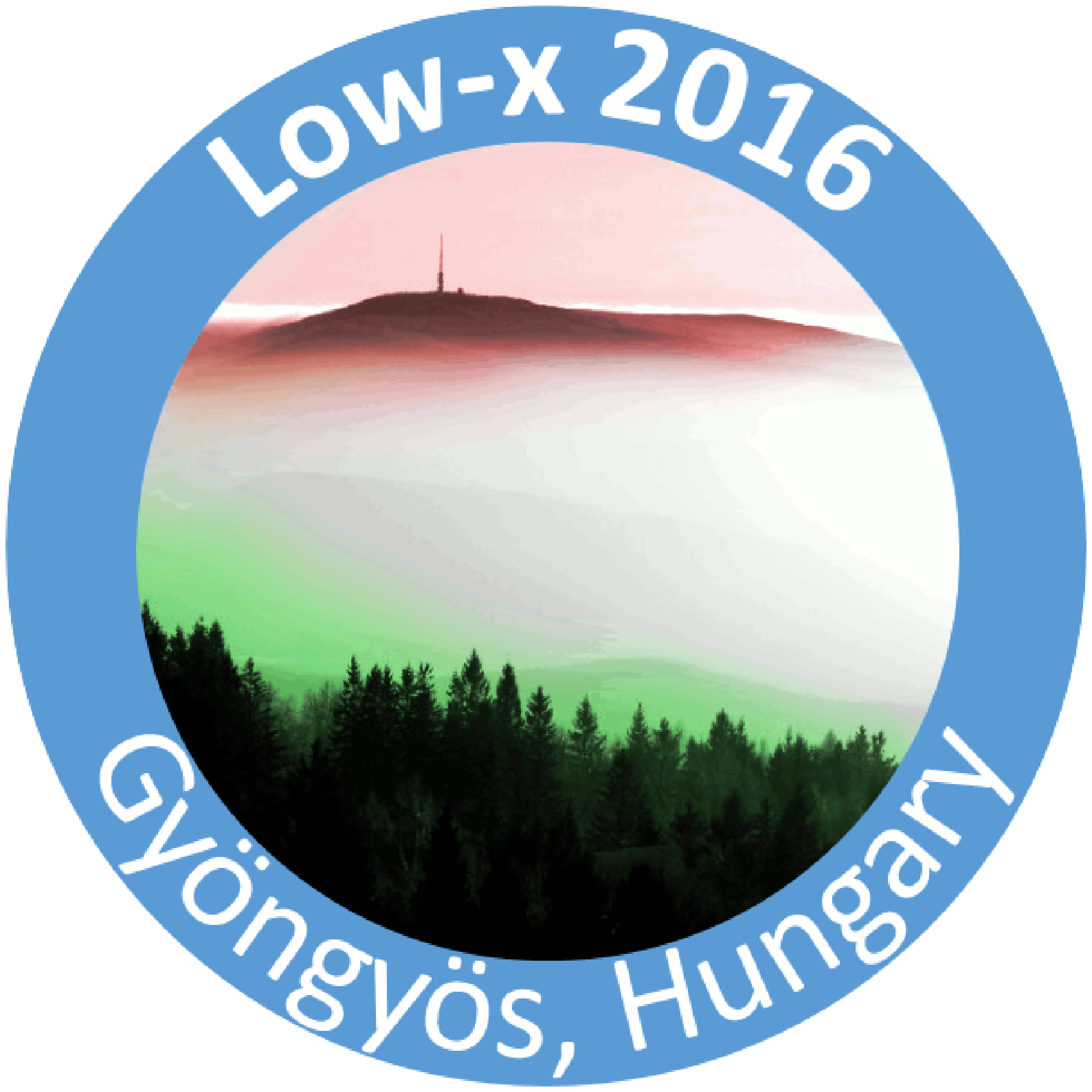}\\[1cm]
Four-jet production in the $k_t$-factorisation
}
\author{{Antoni Szczurek$^{1,2}$, Rafal Maciula$^1$}
\\[1ex]
$^1$Institute of Nuclear Physics PAN, PL-31-342 Krak{\'o}w, Poland\\
$^2$University of Rzesz\'ow, PL-35-959 Rzesz\'ow, Poland\\
}

\begin{document}

\fontfamily{lmss}\selectfont
\maketitle

\begin{abstract}
We discuss the single-parton and double-parton scattering (SPS or DPS) 
effects in four-jet production at the LHC.
The calculations of both single-parton and double-parton scattering
components are done in the high-energy (or $k_{T}$)-factorization
approach. Here we follow our recent developments of relevant methods and
tools. 
The calculations are performed for kinematical situations
relevant for two experimental measurements (ATLAS and CMS) at the LHC.
We compare our results to those reported by the ATLAS and CMS
collaborations for different sets of kinematical cuts.
A special attention is given to 
the optimization of kinematical conditions in order to enhance the
relative contribution of DPS in four-jet sample. 
Several differential distributions are calculated and carefully
discussed in the context of recent and future searches for DPS
effects at the LHC. 
The dependences of the relative DPS amount is studied as a function 
of rapidity of jets, rapidity distance, and various azimuthal 
correlations between jets. The regions with an enhanced 
DPS contribution are identified.
\end{abstract}

\section{Introduction}

So far, complete four-jet production via single-parton scattering 
(SPS) was discussed only within collinear factorization. Results up 
to next-to-leading (NLO) precision can be found in 
\cite{Bern:2011ep,Badger:2012pf}.
Recently we discussed for the first time production of four jets within 
high-energy ($k_T$-)factorization (HEF)
approach with 2 $\to$ 4 subproceses with two
off-shell partons \cite{Kutak:2016mik}.  

Four-jet production seems a natural case to look for hard double-parton 
scattering (DPS) effects (see e.g. Ref.~\cite{Kutak:2016ukc} and 
references therein). Some time ago we analyzed how to find optimal
conditions for the observation and exploration of DPS 
effects in four-jet production \cite{Maciula:2015vza}. In this analysis only 
the leading-order (LO) collinear approach was applied both to single and
double-parton scattering mechanisms.

Very recently, we have performed for the first time a calculation of 
four-jet production for both single-parton and double-parton mechanism 
within $k_T$-factorization \cite{Kutak:2016mik}. 
It was shown that the effective inclusion of higher-order effects leads 
to a substantial damping of the double-scattering contribution 
with respect to the SPS one, especially for symmetric (identical) cuts on the transverse momenta of all jets. 

So far, most practical calculations of DPS contributions were performed 
within the so-called factorized ansatz. 
In this approach, the cross section for DPS is a product 
of the corresponding cross sections for single-parton scatterings (SPS). 
This is a phenomenologically motivated approximation which is not well
under control yet. 
A better formalism exists in principle, but predictions are not easy, 
as they require unknown input(s), \textit{e.g.} double-parton
distributions that should contain informations about
space-configuration, spin, colour or flavour correlations between 
the two partons in one hadron \cite{Diehl:2011yj}. 
These objects are explored to a far lesser extent than the standard 
single PDFs. 
However, the factorized model seems to be a reasonable tool to collect
empirical facts to draw useful conclusions 
about possible identification of the DPS effects
in several processes. 

As discussed in Ref.~\cite{Maciula:2015vza}, jets with low cuts on 
the transverse momenta and a large rapidity separation 
seem more promissing in exploring DPS effects in four-jet 
production.
In the following we shall show our recent results for SPS and DPS
calculations obtained for first time in $k_T$-factorization approach
and concentrate on the study of 
optimal observables to pin down DPS contributions.

\section{A sketch of the theoretical formalism}

The theoretical formalism used to obtain the following predictions was
discussed in detail in \cite{Kutak:2016mik}.
All details related to the scattering amplitudes with off-shell initial 
state partons as well as with the Transverse Momentum Dependent or 
unintegrated parton distribution functions (TMDs) can be found in
our original paper.

Here we only very briefly recall the basic high-energy (or
$k_{T}$)-factorization (HEF) formula for the calculation of 
the inclusive partonic four-jet cross section: 
\begin{eqnarray}
\sigma^B_{4-jets} 
&=& 
\sum_{i,j} \int \frac{dx_1}{x_1}\,\frac{dx_2}{x_2}\, d^2 k_{T1} d^2
k_{T2}\,  
\mathcal{F}_i(x_1,k_{T1},\mu_F)\, \mathcal{F}_j(x_2,k_{T2},\mu_F)
\nonumber \\
&&
\hspace{-25mm}
\times \frac{1}{2 \hat{s}} \prod_{l=i}^4 \frac{d^3 k_l}{(2\pi)^3 2 E_l} 
\Theta_{4-jet} \, (2\pi)^4\, 
\delta\left( x_1P_1 + x_2P_2 + \vec{k}_{T\,1}+ \vec{k}_{T\,2} - \sum_{l=1}^4 k_i \right)\, 
\overline{ \left| \mathcal{M}(i^*,j^* \rightarrow 4\, \text{part.})
\right|^2 } \, . \nonumber \\
\label{kt_cross}
\end{eqnarray}
Above $\mathcal{F}_i(x_k,k_{Tk},\mu_F)$ is the TMD for a given parton
type, $x_k$ are the longitudinal momentum fractions, $\mu_F$ is 
a factorization scale, $\vec{k}_{Tk}$ the parton's transverse momenta.
$\mathcal{M}(i^*,j^* \rightarrow 4\, \text{part.})$ is the gauge
invariant matrix element for $2\rightarrow 4$ particle scattering 
with two initial off-shell partons. They are evaluated numerically 
with the help of the AVHLIB~\cite{Bury:2015dla} Monte Carlo library.
In the calculation, the scales are set to
$\mu_F=\mu_R= \frac{\hat{H}_T}{2} = \frac{1}{2} \sum_{l=1}^4 k_T^l$
\footnote{We use the $\hat{H}_T$ notation to refer to the energies of 
the final state partons, not jets, despite this is obviously the same
in a LO analysis.}.

The so-called pocket formula for DPS cross sections (for a four-parton 
final state) reads:
\begin{equation}
\frac{d \sigma^{B}_{4-jet,DPS}}{d \xi_1 d \xi_2} = 
\frac{m}{\sigma_{eff}} \sum_{i_1,j_1,k_1,l_1;i_2,j_2,k_2,l_2} 
\frac{d \sigma^B(i_1 j_1 \rightarrow k_1 l_1)}{d \xi_1}\, 
\frac{d \sigma^B(i_2 j_2 \rightarrow k_2 l_2)}{d \xi_2} \, ,
\end{equation}
where the $\sigma(a b \rightarrow c d)$ cross sections are obtained by
restricting (\ref{kt_cross}) to a single channel and the symmetry 
factor $m$ is $1/2$ if the two hard scatterings are identical, 
to avoid double counting.
Finally, $\xi_1$ and $\xi_2$ stand for generic kinematical variables for
the first and second scattering, respectively.
The effective cross section $\sigma_{eff}$ can be interpreted as a
measure of correlation in the transverse plane of the two partons inside 
the hadrons, whereas the possible longitudinal correlations are usually 
neglected.
In the numerical calculations here we use $\sigma_{eff}$ = 15 mb that 
is a typical value known from the world systematics 
\cite{Proceedings:2016tff}.

\section{Selected results}

\begin{figure}[!h]
\begin{minipage}{0.47\textwidth}
 \centerline{\includegraphics[width=1.0\textwidth]{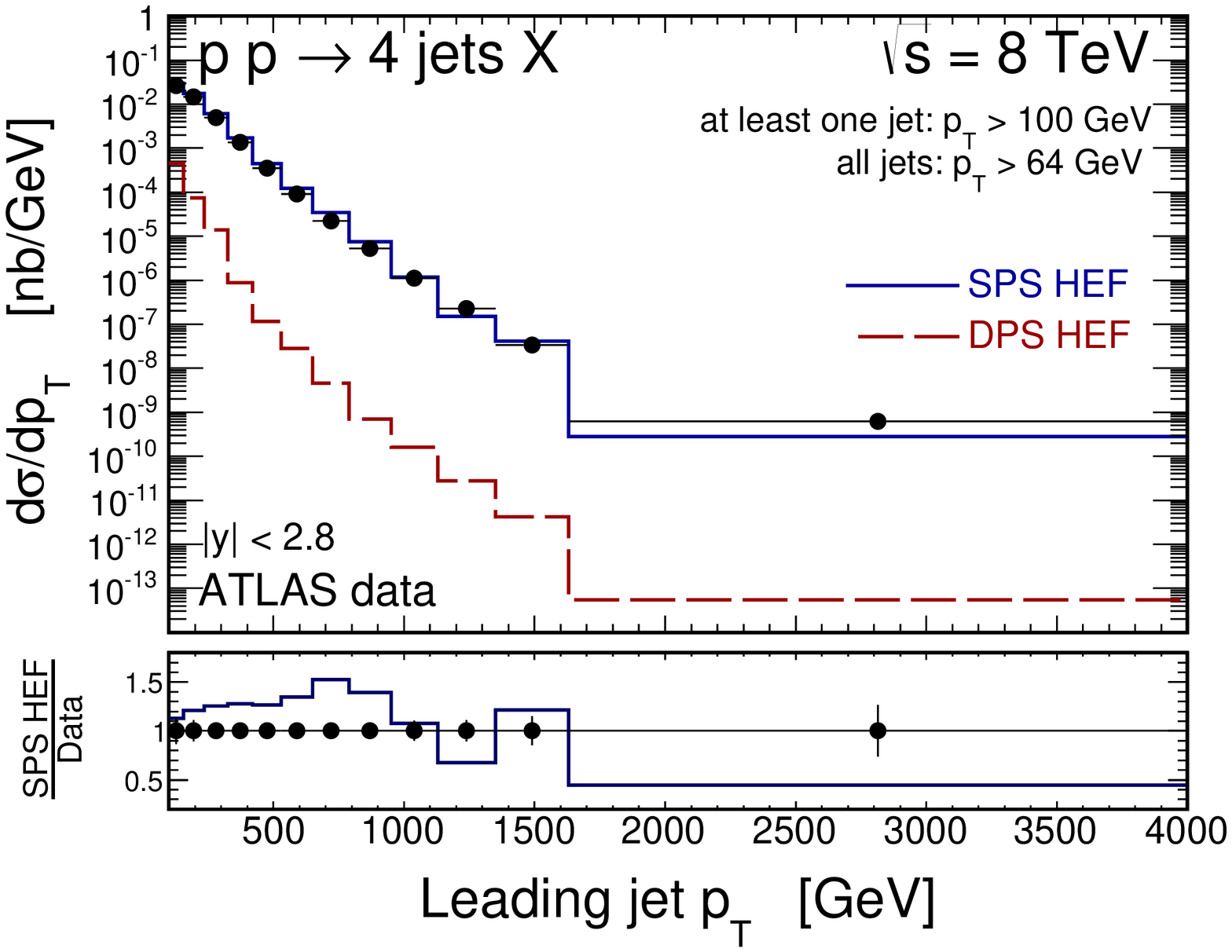}}
\end{minipage}
\hspace{0.5cm}
\begin{minipage}{0.47\textwidth}
 \centerline{\includegraphics[width=1.0\textwidth]{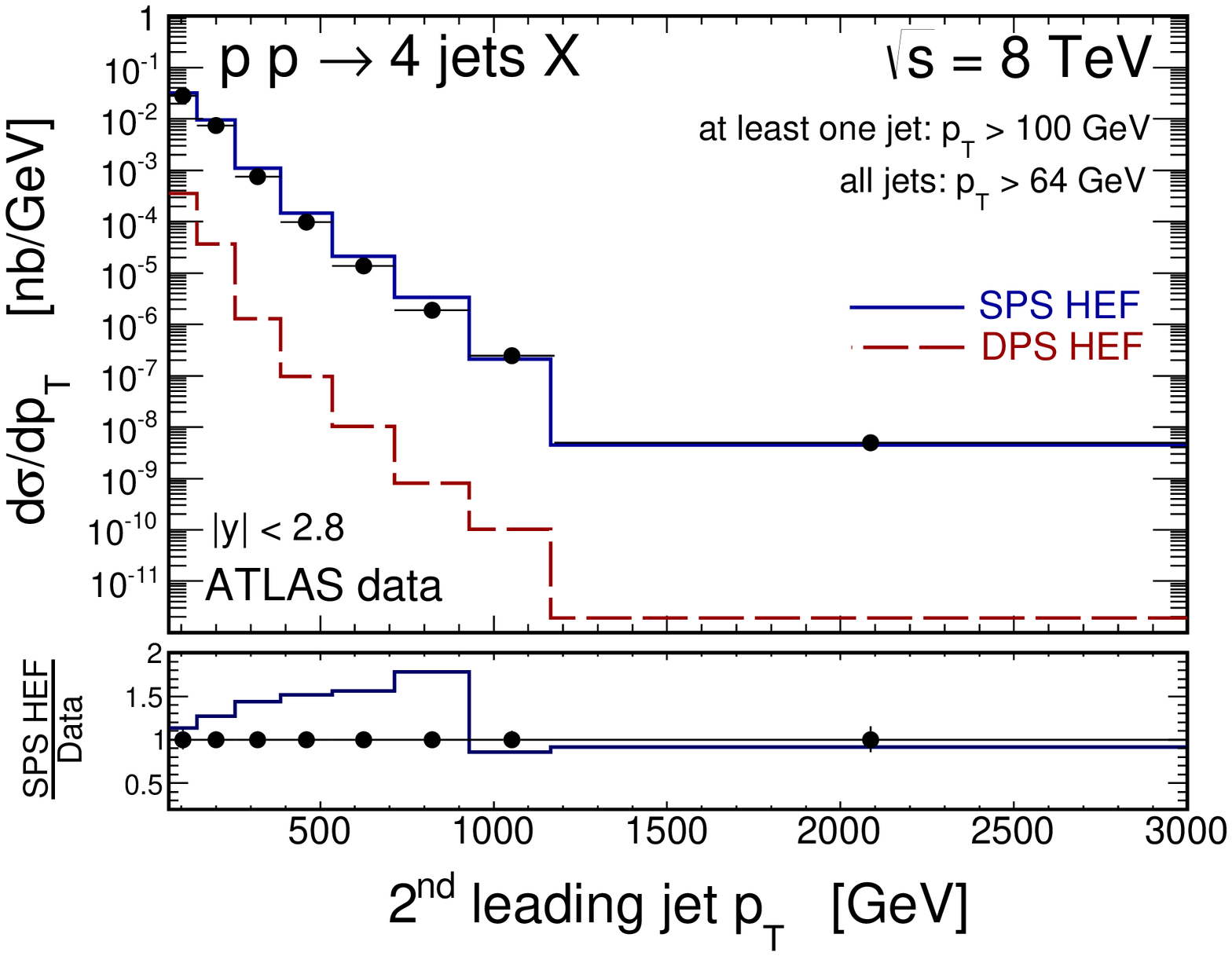}}
\end{minipage}
\caption{
$k_T$-factorization prediction of the differential cross sections
w.r.t. the transverse momenta of the first two leading jets
compared to the ATLAS data \cite{Aad:2015nda}. The LO calculation 
describes the data pretty well in this hard regime in which MPIs are 
irrelevant. In addition we show the ratio of the SPS HEF result 
to the ATLAS data.}
\label{Hard_central_1}
\end{figure}

\begin{figure}[!h]
\begin{minipage}{0.47\textwidth}
 \centerline{\includegraphics[width=1.0\textwidth]{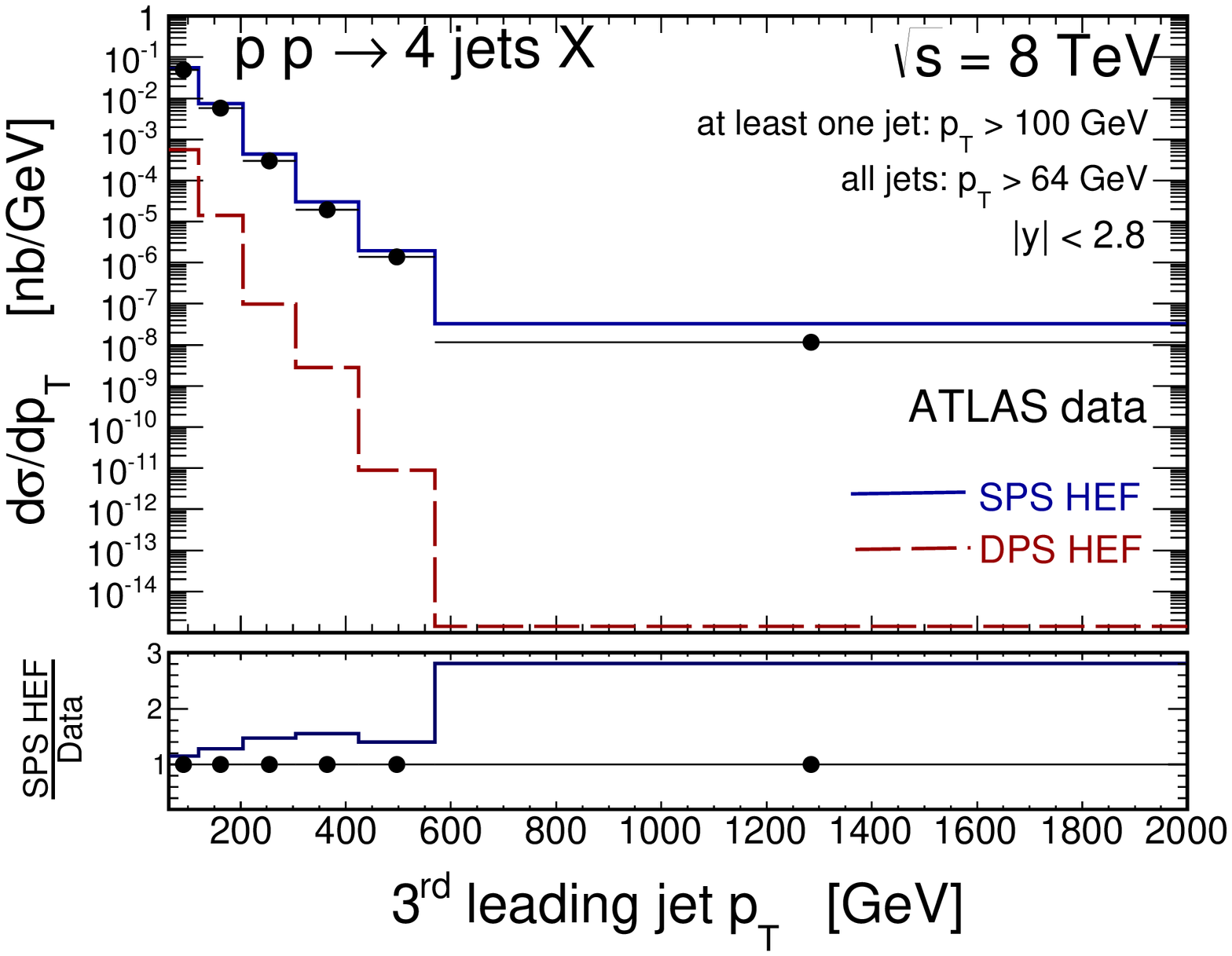}}
\end{minipage}
\hspace{0.5cm}
\begin{minipage}{0.47\textwidth}
 \centerline{\includegraphics[width=1.0\textwidth]{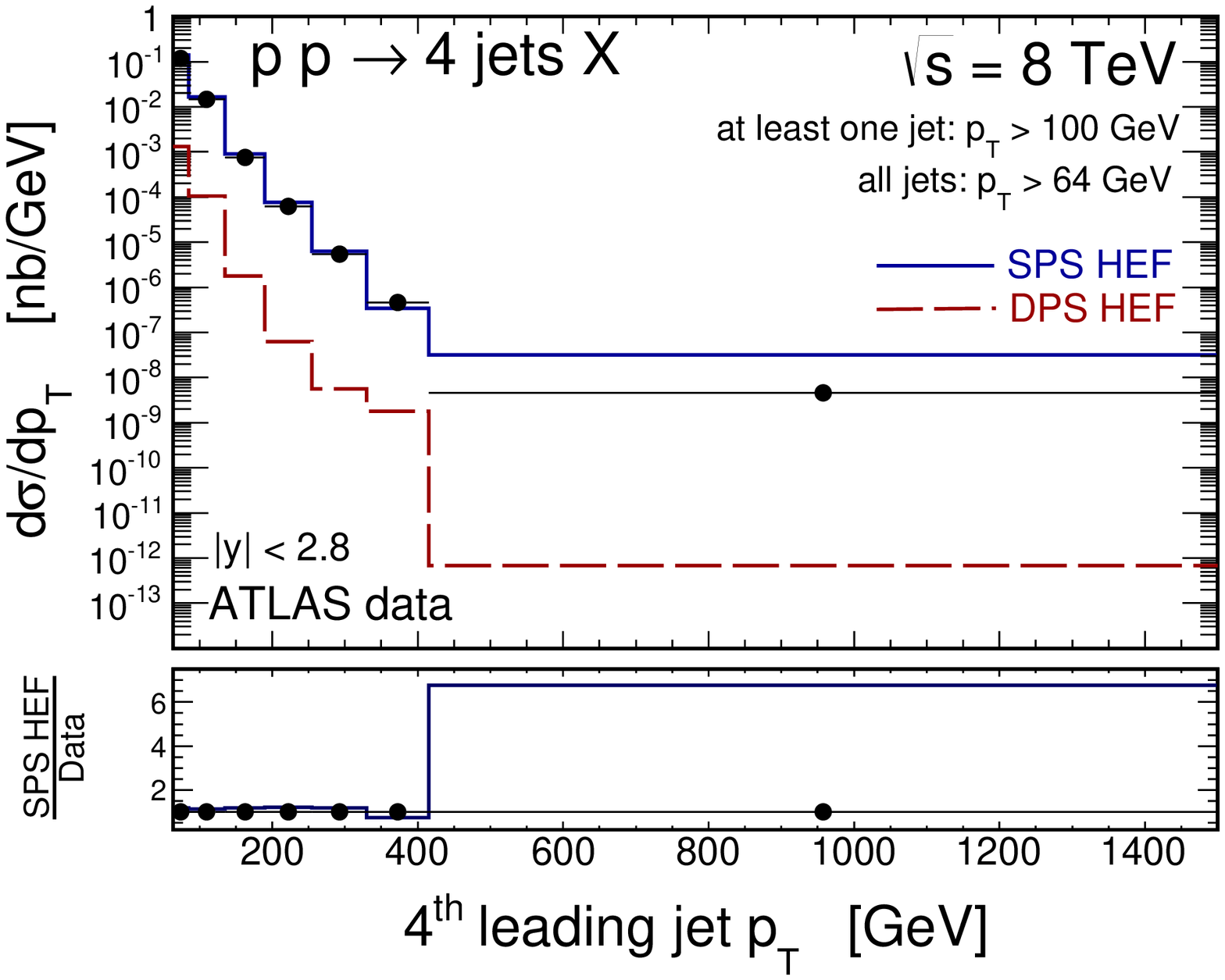}}
\end{minipage}
\caption{
$k_T$-factorization approach prediction of the differential cross
sections w.r.t. the transverse momenta of the 3rd and 4th leading jets
compared to the ATLAS data \cite{Aad:2015nda}. The LO calculation
describes the data pretty well in this hard regime in which MPIs are 
irrelevant. In addition we show the ratio of the SPS HEF result 
to the ATLAS data.}
\label{Hard_central_2}
\end{figure}

First we show some selected examples of the results of the
$k_T$-factorization calculation in Figs.~\ref{Hard_central_1} 
and \ref{Hard_central_2}.
In this calculations we used the KMR unintegrated parton distributions.
The prediction is consistent with the ATLAS data for all the $p_T$ 
distributions. 

%

Not only transverse momentum dependence is interesting.
The CMS collaboration extracted for instance a more complicated 
observables \cite{Chatrchyan:2013qza}.
One of them, which involves all four jets in the final state,
is the $\Delta S$ variable, defined in Ref.~\cite{Chatrchyan:2013qza} 
as the angle between pairs of the harder and the softer jets,
\begin{equation}
\Delta S = \arccos \left( 
\frac{\vec{p}_T(j^{\text{hard}}_1,j^{\text{hard}}_2) \cdot \vec{p}_T(j^{\text{soft}}_1,j^{\text{soft}}_2)}{|\vec{p}_T(j^{\text{hard}}_1,j^{\text{hard}}_2)|\cdot|\vec{p}_T(j^{\text{soft}}_1,j^{\text{soft}}_2)|}  
\right) \, ,
\end{equation}
where $\vec{p}_T(j_i,j_k)$ stands for the sum of the transverse momenta of the two jets in arguments.

\begin{figure}[h]
\begin{center}
\begin{minipage}{0.47\textwidth}
 \centerline{\includegraphics[width=1.0\textwidth]{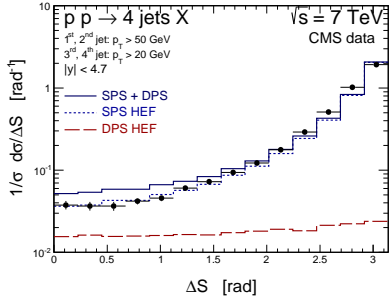}}
\end{minipage}
\end{center}
\caption{
Comparison of the HEF predictions to the CMS data for $\Delta S$
spectrum. 
}
\label{fig:CMS_DS}
\end{figure}

In Fig.~ \ref{fig:CMS_DS} we present our HEF prediction for
the normalized to unity distribution in the $\Delta S$ variable. 
Our HEF result approximately agrees with the
experimental $\Delta S$ distribution. 
In contrast, the LO collinear approach leads to 
$\Delta S$ = 0, i.e. a Kronecker-delta peak at $\Delta S$ = 0 for 
the distribution in $\Delta S$.

%
%

\begin{figure}[!h]
\begin{minipage}{0.47\textwidth}
 \centerline{\includegraphics[width=1.0\textwidth]{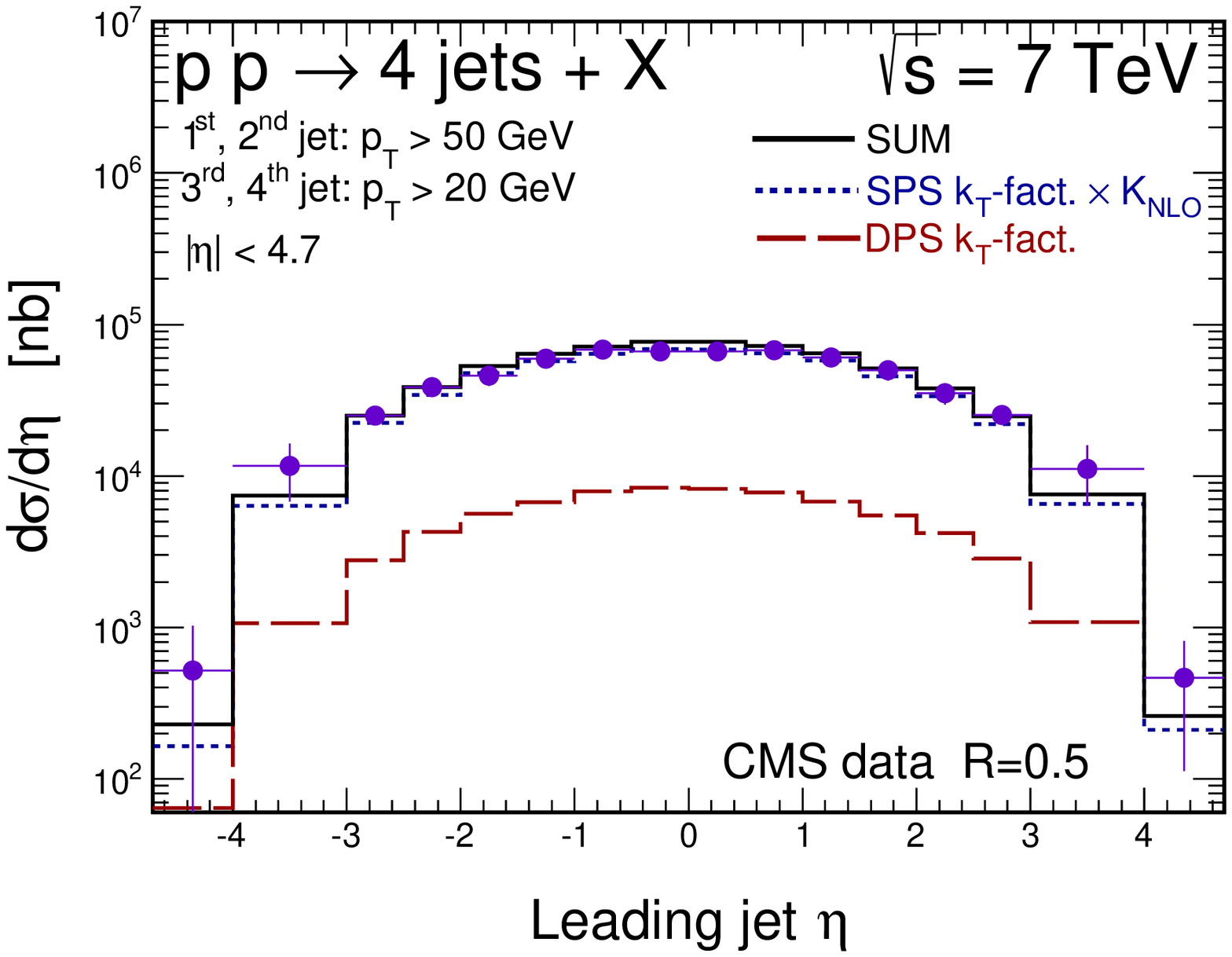}}
\end{minipage}
\hspace{0.5cm}
\begin{minipage}{0.47\textwidth}
 \centerline{\includegraphics[width=1.0\textwidth]{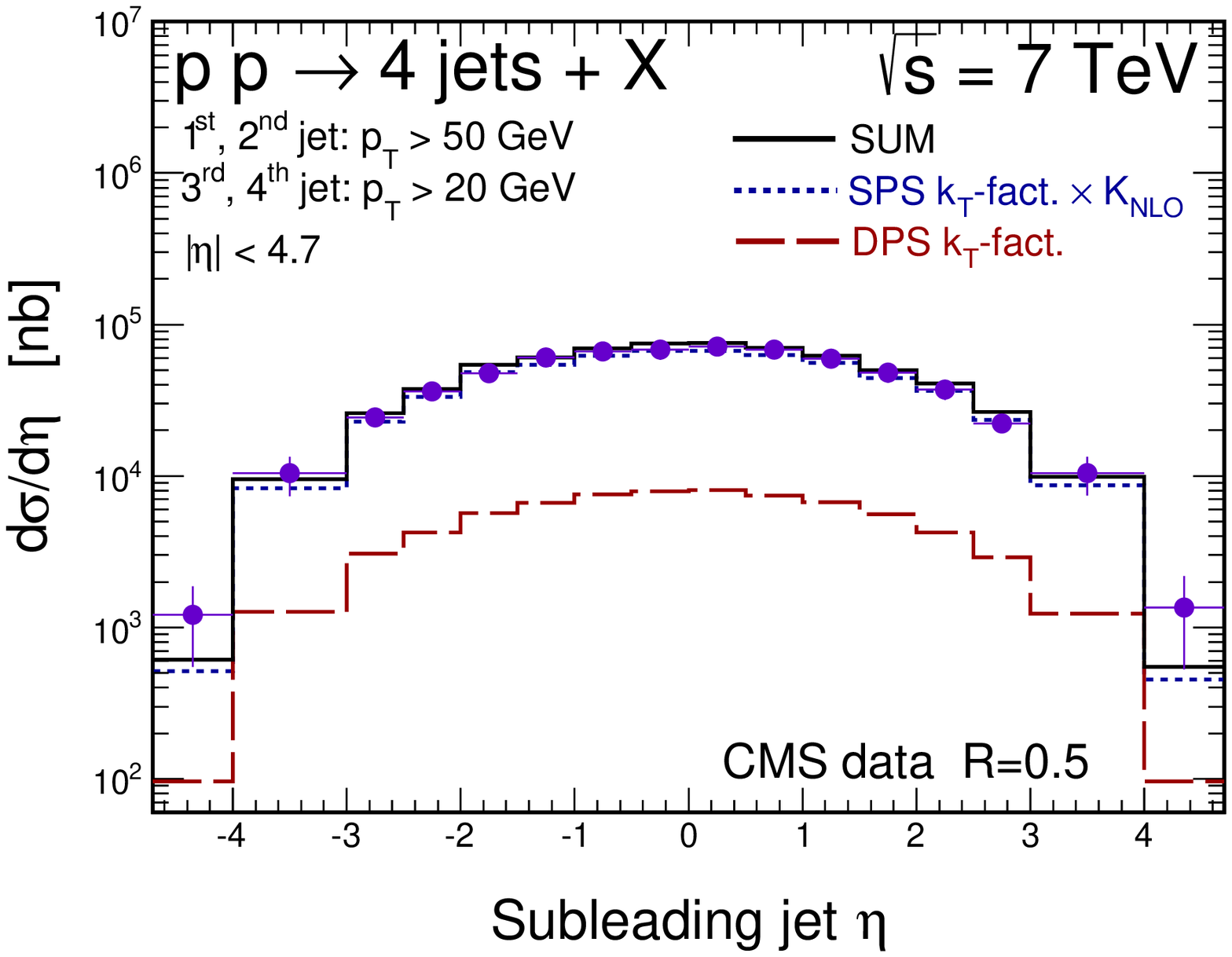}}
\end{minipage}
   \caption{
\small Rapidity distribution of the leading and subleading jets.
The SPS contribution is shown by the dotted line while the DPS contribution by the dashed line.
 }
 \label{fig:CMS_y_distributions}
\end{figure}

Now we wish to show a comparison of our numerical predictions with
existing experimental data for relatively low cuts on jet transverse momenta.
In this context, the CMS experimental multi-jet analysis
\cite{Chatrchyan:2013qza} is the most 
relevant as it uses sufficiently soft cuts on the jet transverse momenta.
The cuts are in this case $|p_T| > 50$ GeV for the two hardest jets 
and $|p_T| > 20$ GeV for the third and fourth ones; the rapidity region
is defined by $|\eta| < 4.7$ and the constraint on the jet cone radius
parameter is $\Delta R >0.5$.
The overall situation is shown in Fig.~\ref{fig:CMS_y_distributions}, 
where we plot rapidity distributions for leading and subleading jets
ordered by their $p_{T}$'s.

The $k_T$-factorization approach includes higher-order corrections 
through the resummation in the TMDs. However,
within this framework fixed-order loop effects are not taken into account.
Therefore, we allow for a $K$-factor for the calculation of the SPS component.
The NLO $K$-factors are known to be smaller than unity
for 3- and 4-jet production in the collinear approximation case \cite{Bern:2011ep}. 
To describe the CMS data, we also need $K$-factors smaller than unity
for the SPS contributions, as expected.
In contrast to the 4-jet case, the NLO predictions for the 2-jet inclusive cross section 
are further away from the measured value than the LO ones
\cite{Bern:2011ep}. The 2-jet $K$-factor is known to be about $1.2$,
and it enters squared in the case of the DPS calculations. 
However, in our calculations we ignored the relatively small $K$-factors for 
the DPS contribution.

\begin{figure}[!h]
\begin{minipage}{0.47\textwidth}
 \centerline{\includegraphics[width=1.0\textwidth]{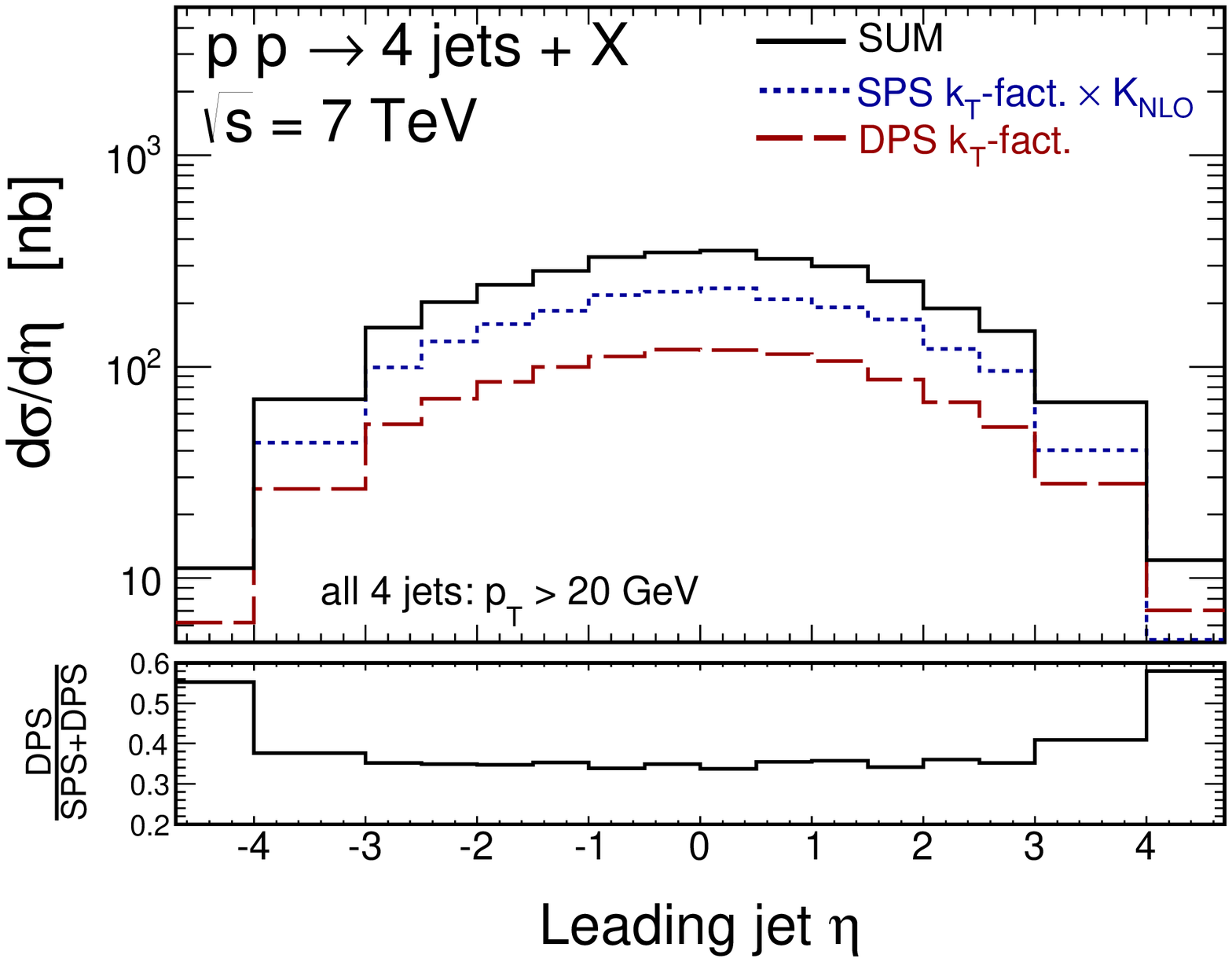}}
\end{minipage}
\hspace{0.5cm}
\begin{minipage}{0.47\textwidth}
 \centerline{\includegraphics[width=1.0\textwidth]{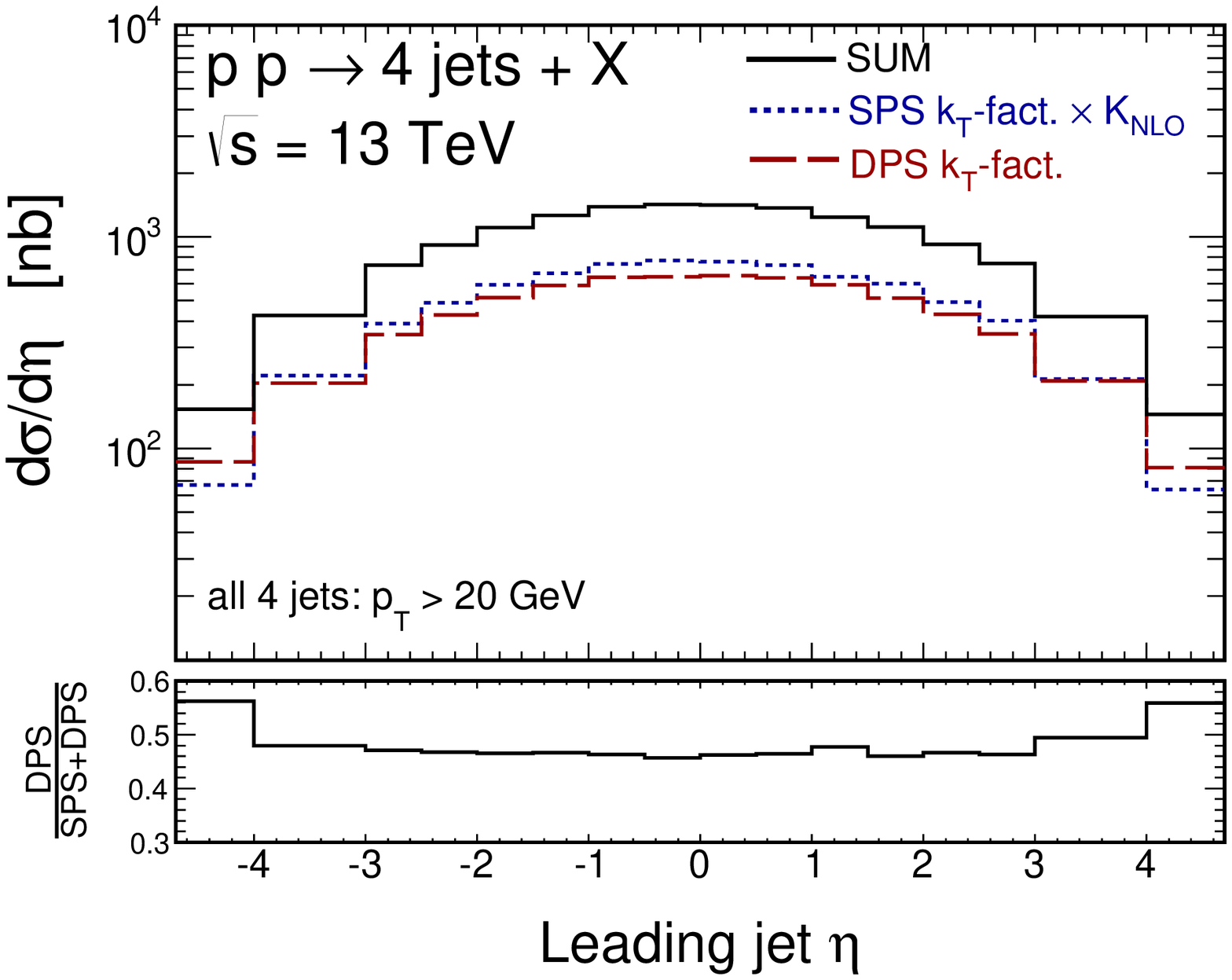}}
\end{minipage}
   \caption{
\small Rapidity distribution of leading jet for $\sqrt{s}$ = 7 TeV
(left column) and $\sqrt{s}$ = 13 TeV (right column) for
the symmetric cuts. The SPS contribution
is shown by the dotted line while the DPS contribution by the dashed line.
The relative contribution of DPS is shown in the extra lower panels.
 }
 \label{fig:dsig_dy_symmetric_20GeV}
\end{figure}

In Refs.\cite{Maciula:2015vza,Kutak:2016ukc} we introduced a set 
of observables that we find particularly convenient to
identify DPS effects in four-jet production.
Here we present results for completely symmetric cuts, $p_T > 20$ GeV, 
for all the four leading jets.
The cuts on rapidity and jet radius parameter are the same as for 
the CMS case.
In Fig.~\ref{fig:dsig_dy_symmetric_20GeV} we show our predictions
for the rapidity distributions. In contrast to the previous case 
(Fig.~\ref{fig:CMS_y_distributions}), where harder cuts on the two
hardest jets were used, the shapes of the SPS and DPS rapidity distributions 
are rather similar. There is only a small relative enhancement of the
DPS contribution for larger jet rapidities.

As it was proposed first in Ref.~\cite{Maciula:2014pla} in the context 
of Mueller-Navelet jet production, and then
repeated in Ref.~\cite{Maciula:2015vza} for four-jet studies in 
the LO collinear approach, there are two potentially
useful observables for DPS effects, such as the maximum rapidity distance 
\begin{equation}
\Delta \text{Y} \equiv max_{\substack{i,j \in\{1,2,3,4\}\\i \neq j }} |\eta_i-\eta_j |
\end{equation}
and the azimuthal correlations between the jets which are most remote 
in rapidity
\begin{equation}
\varphi_{jj} \equiv | \varphi_i -\varphi_j |  \, , \quad \text{for}  \quad |\eta_i - \eta_j | = \Delta \text{Y} \, .
\end{equation}

\begin{figure}[!h]
\begin{minipage}{0.47\textwidth}
 \centerline{\includegraphics[width=1.0\textwidth]{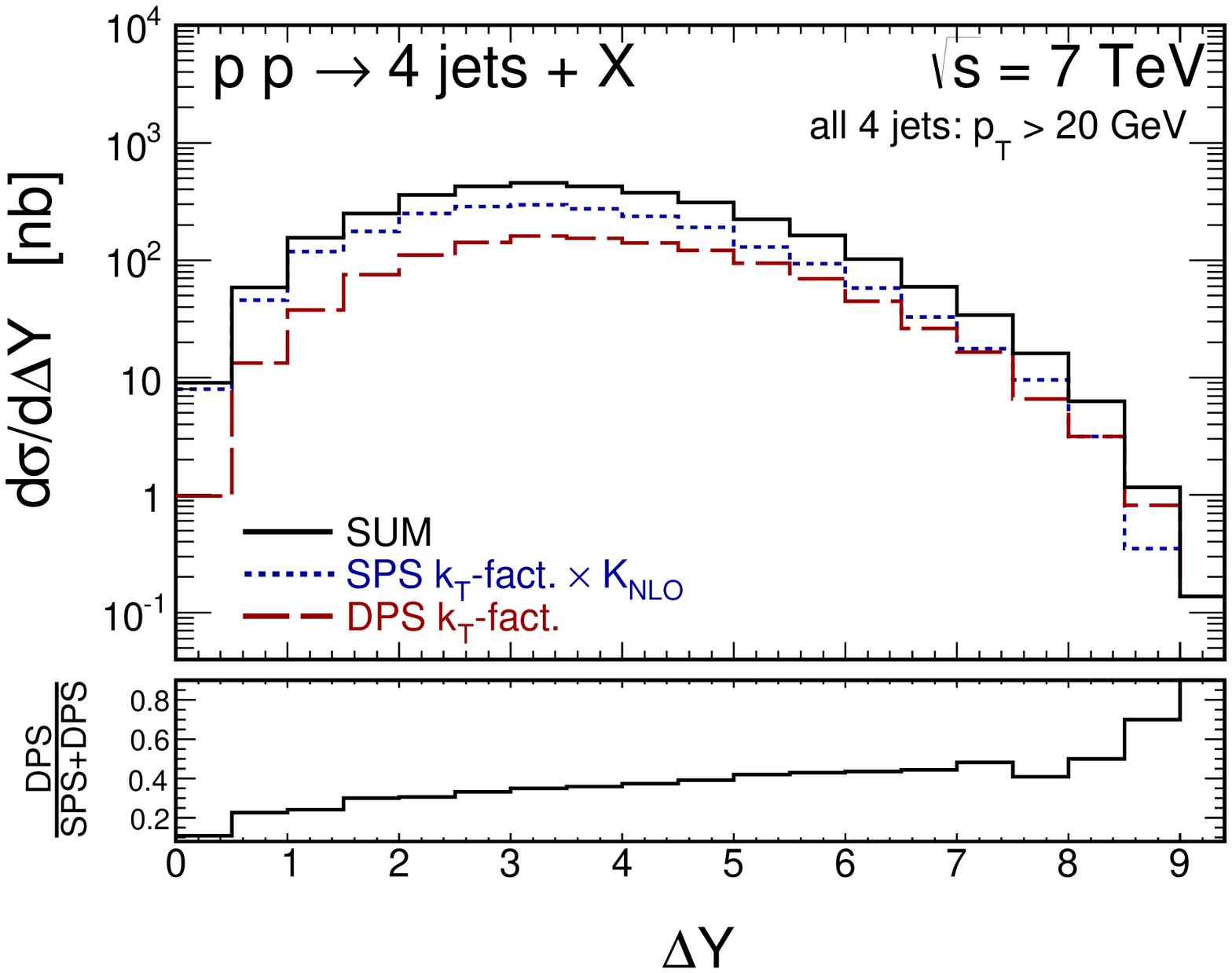}}
\end{minipage}
\hspace{0.5cm}
\begin{minipage}{0.47\textwidth}
 \centerline{\includegraphics[width=1.0\textwidth]{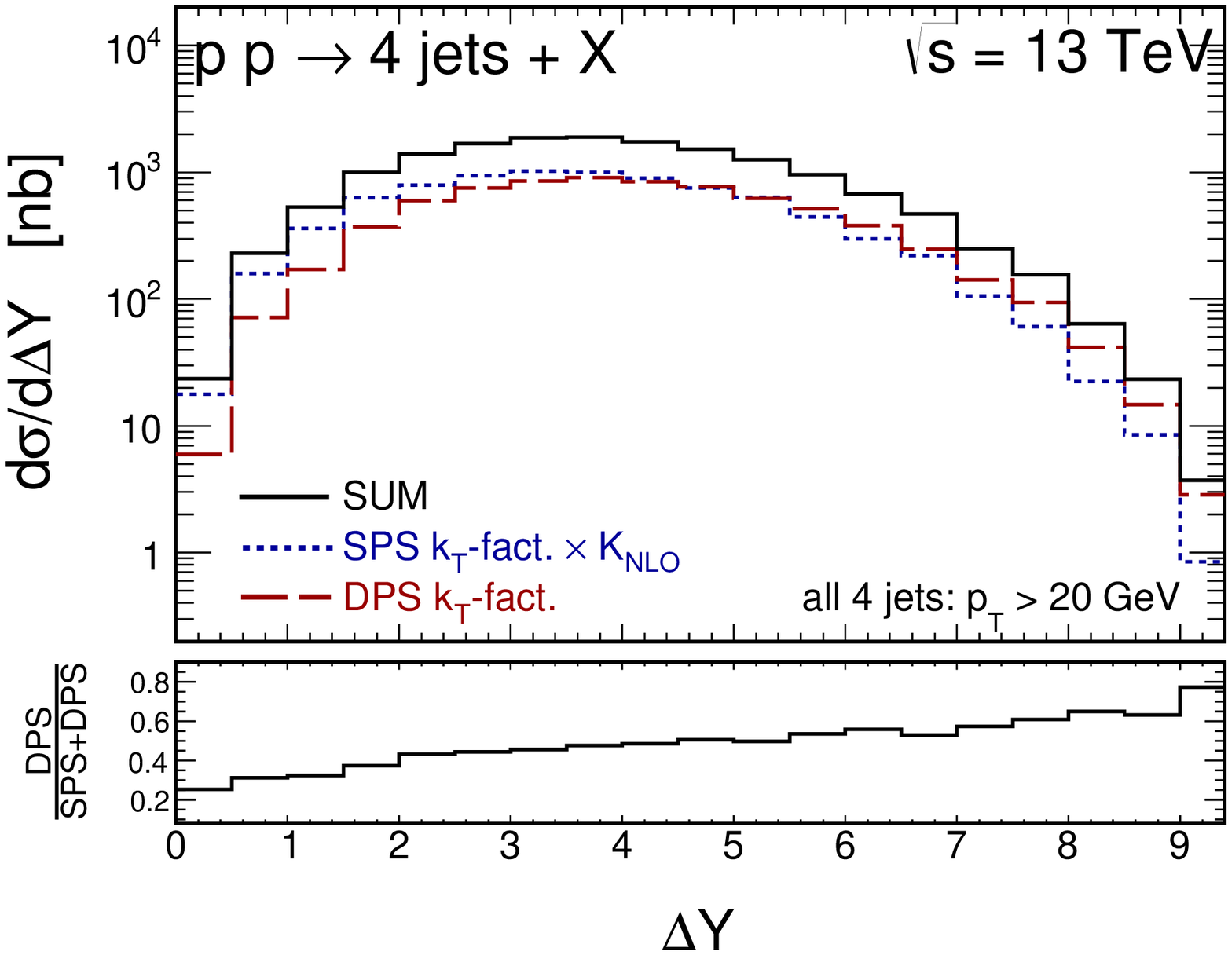}}
\end{minipage}
   \caption{
\small Distribution in rapidity distance between the most remote jets for 
the symmetric cut with $p_T >$ 20 GeV 
for $\sqrt{s}$ = 7 TeV (left) and $\sqrt{s}$ = 13 TeV (right).
The SPS contribution is shown by the dotted line while 
the DPS contribution by the dashed line.
The relative contribution of DPS is shown in the extra lower panels.
 }
 \label{fig:dsig_dydiff}
\end{figure}

One can see in Fig.~\ref{fig:dsig_dydiff} that the relative DPS
contribution increases with $\Delta \text{Y}$ which, for 
the CMS collaboration is up to 9.4.
At $\sqrt{s}$ = 13 TeV the DPS component dominates over the SPS
contribution for $\Delta \text{Y} > 6$.
A potential failure of the SPS contribution to describe such a plot 
in this region would be a signal of the presence of a sizable DPS contribution.

\begin{figure}[!h]
\begin{minipage}{0.47\textwidth}
 \centerline{\includegraphics[width=1.0\textwidth]{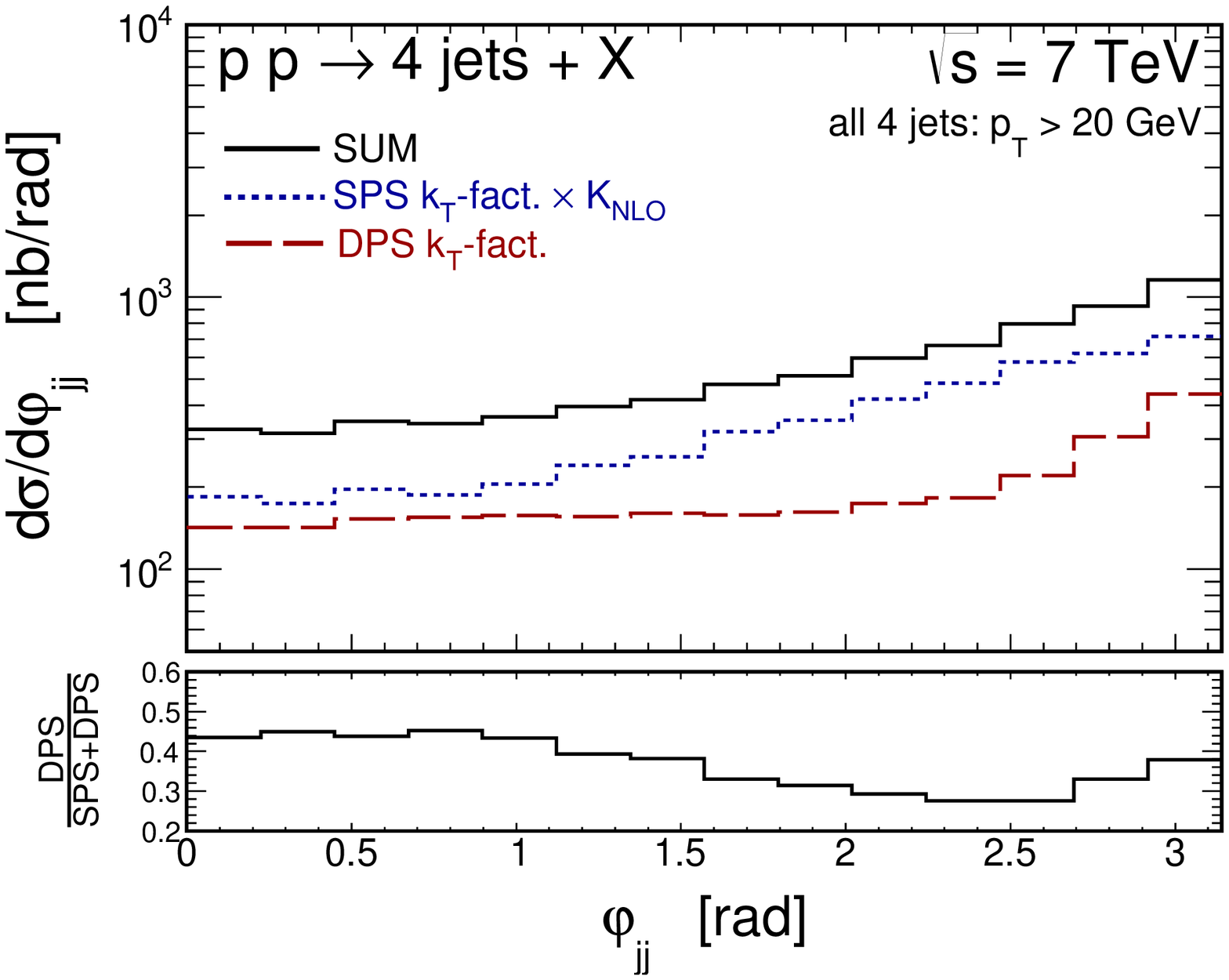}}
\end{minipage}
\hspace{0.5cm}
\begin{minipage}{0.47\textwidth}
 \centerline{\includegraphics[width=1.0\textwidth]{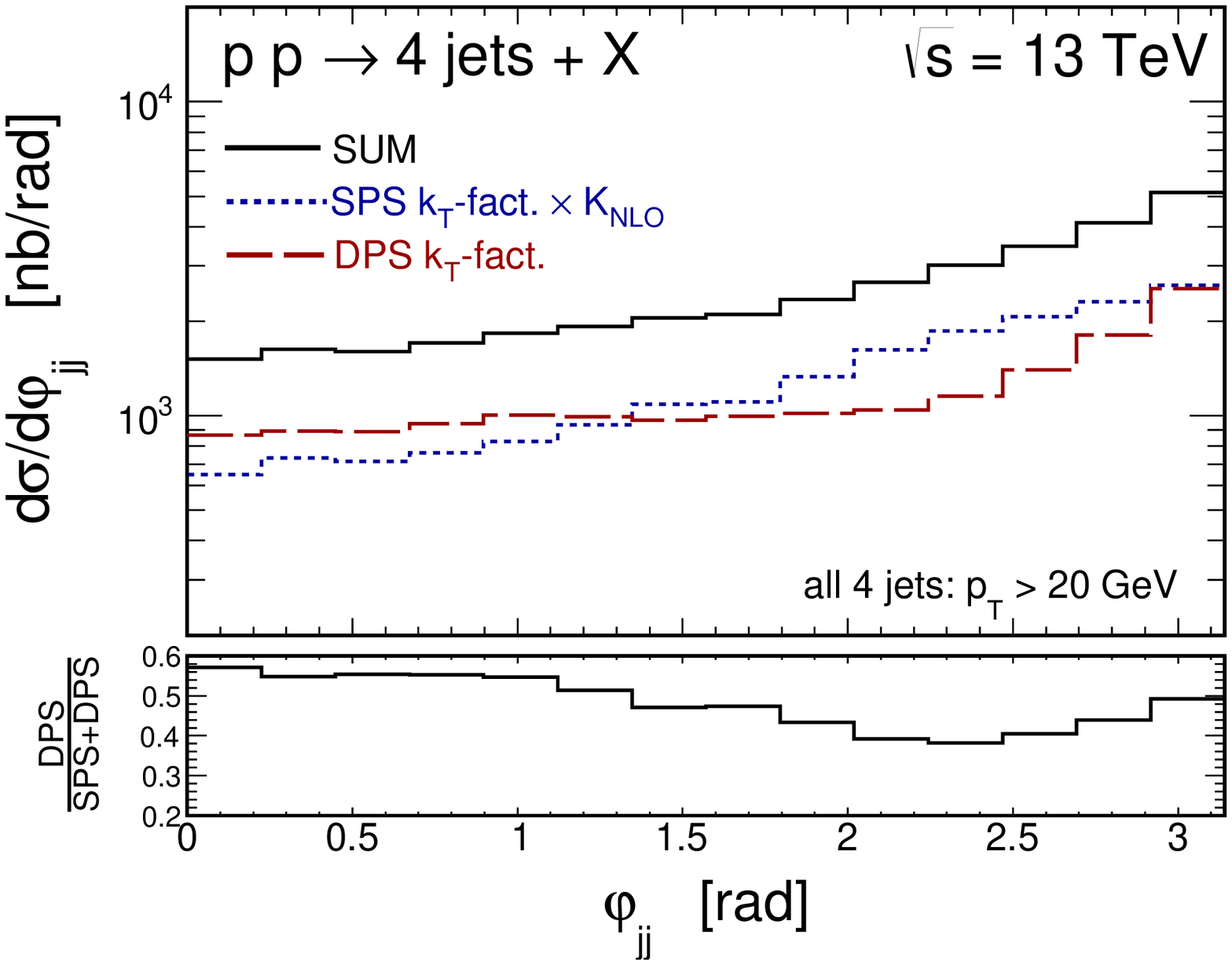}}
\end{minipage}
   \caption{
\small Distribution in relative azimuthal angle between the most remote jets for 
the symmetric cut with $p_T >$ 20 GeV 
for $\sqrt{s}$ = 7 TeV (left) and $\sqrt{s}$ = 13 TeV (right).
The SPS contribution is shown by the dotted line while 
the DPS contribution by the dashed line.
The relative contribution of DPS is shown in the extra lower panels.
 }
 \label{fig:dsig_dphijj}
\end{figure}

Figure~\ref{fig:dsig_dphijj} shows azimuthal correlations between the jets most remote in rapidity. 
While at $\sqrt{s}$~=~7~TeV the SPS contribution is always larger than the DPS one, 
at $\sqrt{s}$ = 13 TeV the DPS component dominates over the SPS contribution for $\varphi_{jj} < \pi/2$. 

\begin{figure}[!h]
\begin{minipage}{0.47\textwidth}
 \centerline{\includegraphics[width=1.0\textwidth]{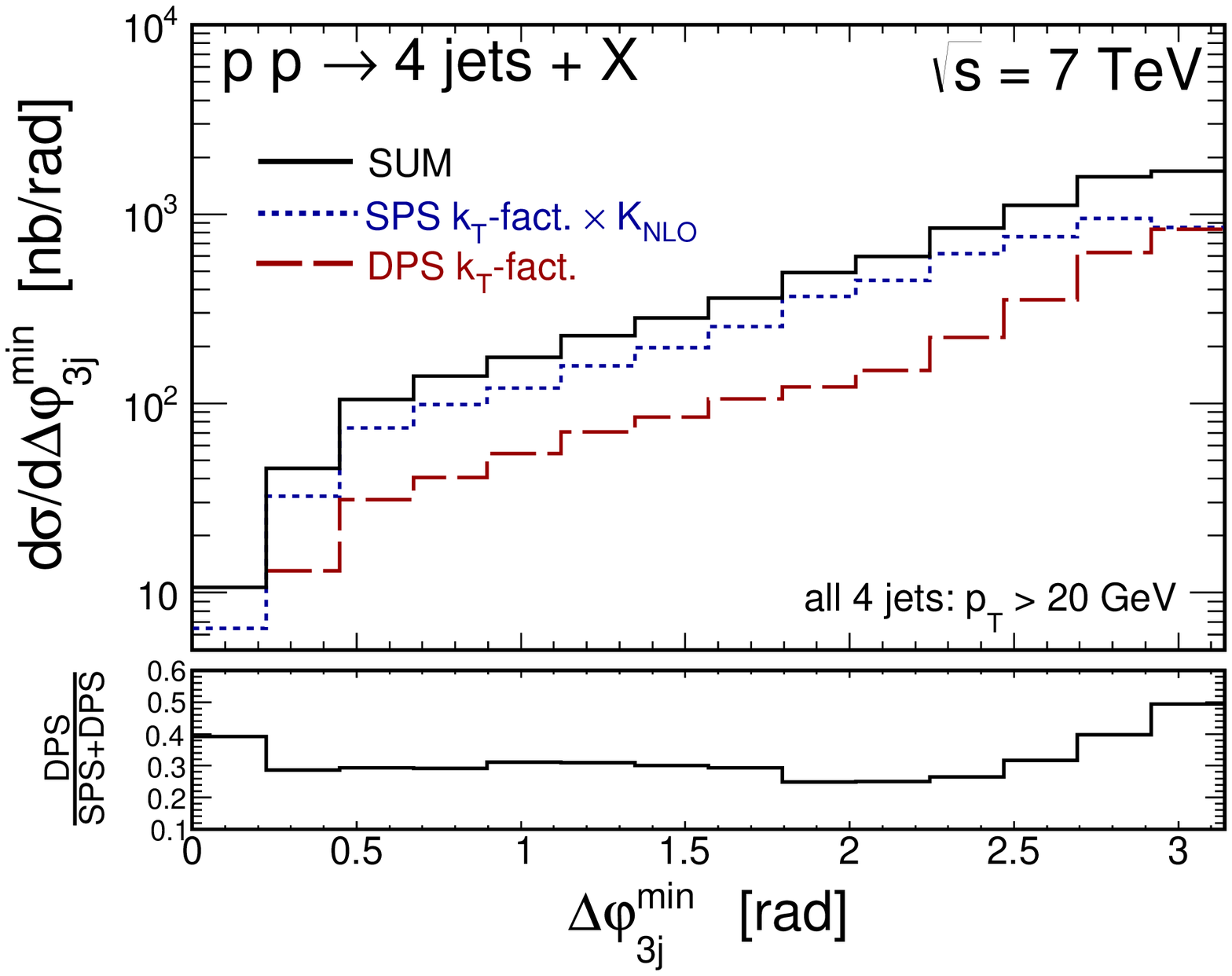}}
\end{minipage}
\hspace{0.5cm}
\begin{minipage}{0.47\textwidth}
 \centerline{\includegraphics[width=1.0\textwidth]{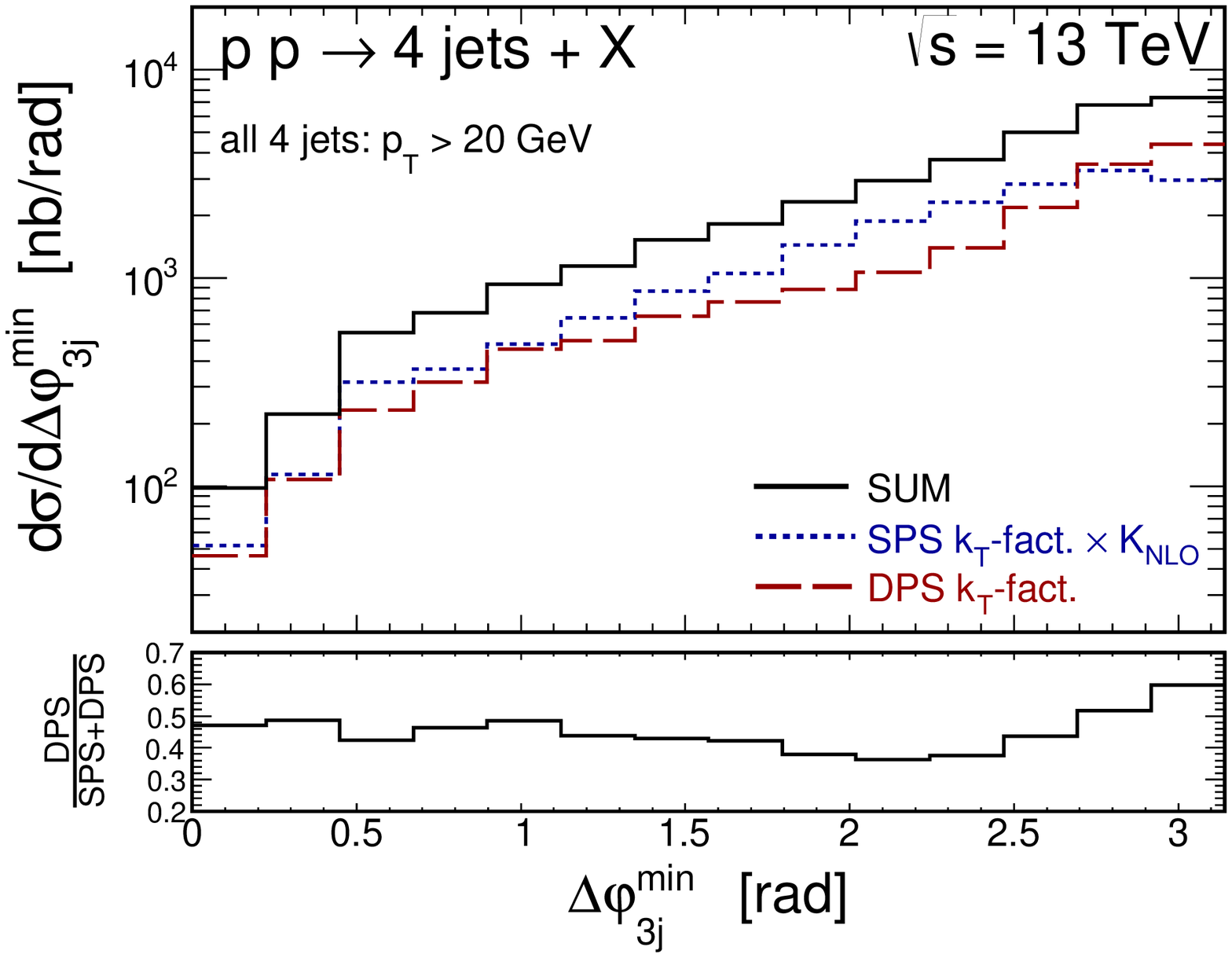}}
\end{minipage}
   \caption{
\small Distribution in $\Delta \varphi_{3j}^{min}$ angle for 
the symmetric cut with $p_T >$ 20 GeV 
for $\sqrt{s}$ = 7 TeV (left) and $\sqrt{s}$ = 13 TeV (right).
The SPS contribution is shown by the dotted line while 
the DPS contribution by the dashed line.
The relative contribution of DPS is shown in the extra lower panels.
 }
 \label{fig:dsig_dphi_3j}
\end{figure}

We also find that another variable, introduced in the high transverse
momenta analysis of four jets production discussed 
in Ref.~\cite{Aad:2015nda}, can be very interesting for 
the examination of the DPS effects:
\begin{equation}
\Delta \varphi_{3j}^{min} \equiv min_{\substack{i,j,k \in\{1,2,3,4\}\\i 
\neq j \neq k}}\left(|\varphi_i - \varphi_j|+| \varphi_j - \varphi_k|\right) \, .
\label{DeltaPhiMin}
\end{equation}
As three out of four azimuthal angles are always entering in 
(\ref{DeltaPhiMin}), configurations with one jet recoiling against 
the other three are necessarily characterised by lower
values of $\Delta \varphi_{3j}^{min}$ with respect to 
the two-against-two topology;
the minimum, in fact, will be obtained in the first case for $i,j,k$ 
denoting the three jets in the same hemisphere, 
whereas no such a case is possible for the second configuration.
Obviously, the first case would be allowed only by SPS in 
a collinear tree-level framework,
whereas the second would be enhanced by DPS. 
In the $k_T$-factorization approach, 
this situation is smeared out by the presence of transverse momenta of 
the initial state partons. 
For our unintegrated parton distributions, the corresponding
distributions are shown in Fig.~\ref{fig:dsig_dphi_3j}.
We do not see such obvious effects in the case of the $k_{T}$-factorization.

\section{Conclusions}

We have presented our recent results for four-jet production
obtained for the first time within $k_T$-factorization approach.
The calculation of the SPS contribution is a technical
achievment. So far only production of the $c \bar c c \bar c$ final
state (also of the 2 $\to$ 4 type) was discussed in the literature. 

We have found that both collinear and the ($k_T$-)factorization 
approaches describe the data for hard central cuts, relevant for 
the ATLAS experiment, reasonably well when using the KMR TMDs. 
For the harder cuts we get both
normalization and shape of the transverse momentum distributions.
We nicely describe also CMS distribution for a special variable $\Delta S$.

In this presentation we have discussed also how to look at the DPS effects
and how to maximize their role in four jet production. 
We found that, for sufficiently small cuts on the transverse momenta, 
DPS effects are enhanced relative to the SPS contribution:
when rapidities of jets  are large, for large rapidity distances between
the most remote jets, for small azimuthal angles between the two jets
most remote in rapidity and/or for large values of 
the $\Delta\varphi_{3j}^{min}$ variable. 
For more details we refer the interested reader to our regular 
article \cite{Kutak:2016ukc}.

\section*{Acknowledgments}
The work presented here was done in collaboration with
Krzysztof Kutak, Mirko Serino and Andreas van Hameren.
We are indebted to them for a friutfull collaboration.
This work was partially supported by DEC-2014/15/B/ST2/02528. 

\end{document}